\documentclass{article}

\usepackage[english]{babel}

\usepackage[letterpaper,top=3cm,bottom=3cm,left=3cm,right=3cm,marginparwidth=1.75cm]{geometry}
 \usepackage{setspace}
\usepackage{booktabs}

\usepackage{amsmath}
\usepackage{graphicx}
\usepackage{natbib}

\usepackage[colorlinks=true, allcolors=blue]{hyperref}

\title{Promises and Pitfalls of a New Early Warning System for Gentrification in Buffalo, NY }
\author{Jan Voltaire Vergara,\textsuperscript{a} Maria Y. Rodriguez,\textsuperscript{b} Ehren Dohler,\textsuperscript{c}\\Jonathan Phillips,\textsuperscript{c} Melissa Villodas,\textsuperscript{c} Amy Blank Wilson,\textsuperscript{b} Kenneth Joseph\textsuperscript{a,1} \\ 
{\small \textsuperscript{a} Department of Computer Science and Engineering, University at Buffalo} \\ 
{\small \textsuperscript{b} School of Social Work, University at Buffalo} \\
{\small \textsuperscript{c} School of Social Work, University of North Carolina At Chapel Hill} \\
{\small \textsuperscript{1} Corresponding Author: kjoseph@buffalo.edu}}
\date{}

\begin{document}
\maketitle

\begin{abstract}
Gentrification and its resultant displacement are one of the many ‘wicked problems’ of social policy. The study of gentrification and displacement spans half a century, concerns a variety of spatial, temporal, and social contexts, and describes socio-political processes of across the globe and throughout history. One current iteration of this field of inquiry are efforts to identify "early indicators" of gentrification and/or displacement, or the creation of ``early warning systems” (EWS). The current work adds to scholarship on the utility of developing an EWS by examining the methodological considerations required for such systems to serve a justice-oriented preventative role. 
\end{abstract}

\section{Introduction}

The literature on gentrification and displacement spans well over a half century. Most scholarship traces the origins of gentrification research back to \citeauthor{glass}' \citeyear{glass} work on gentrifying neighborhoods in London, and work on displacement to the work of \citet{grier1980urban}.  Research on both topics spans myriad methodologies, ranging from deep ethnography to the use of "big data" and machine learning \citep{hwangDivergentPathwaysGentrification2014a,readesUnderstandingUrbanGentrification2019}. Scholarship also spans myriad spatial, temporal, and social contexts, describing the socio-political processes of gentrification and displacement across the globe and throughout history.

Despite the range of methods and contexts studied, however, there are common threads running through existing work. One point of emphasis has been on the various negative effects of gentrification and displacement on city residents. Because of these negative effects, a second common theme is one of finding "early indicators" of gentrification and/or displacement. By identifying such indicators, we can potentially mitigate negative impacts via interventions, such as enacting policies like rent-control. Especially within city governments, considerable effort has therefore been put into the development of Early Warning Systems (EWS), tools that can be used to detect gentrification and/or displacement in their early stages \cite{chappleUseNeighborhoodEarly2016}. These efforts have met mixed success, with some work suggesting the promise of such systems \citep{readesUnderstandingUrbanGentrification2019} and others emphasizing caution in their application, particularly across contexts \citep{preisMappingGentrificationDisplacement2020}. 

These promises and pitfalls of EWS for gentrification/displacement arise throughout a complex pipeline of development that involves four deeply inter-related challenges: \emph{data selection}, \emph{outcome operationalization}, \emph{model construction},  and \emph{model  evaluation}.\footnote{These are roughly the same elements noted by \citet{readesUnderstandingUrbanGentrification2019}, although ours are more aligned with typical models of the ``machine learning pipeline'' (e.g. \url{https://nlpers.blogspot.com/2016/08/debugging-machine-learning.html})} 

With respect to data selection, most existing EWS in the U.S. use traditional, widely-available data sources, in particular, annual population estimates from the U.S. census \cite{zukGentrificationDisplacementRole2018}. However, scholars are increasingly turning to other forms of data \citep{bartonExplorationImportanceStrategy2016}, such as anonymized surveys \citep{carlsonMeasuringDisplacementAssessing2021} and indicators available through other forms of administrative data \citep{readesUnderstandingUrbanGentrification2019}, to address limitations of census data. 

With respect to outcome operationalization, EWS developers face the decades-old question of how to quantify gentrification and/or displacement \citep{zukGentrificationDisplacementRole2018,bradwaylaskaInnerCityReinvestmentNeighborhood1982}. As noted by \citet{eastonMeasuringMappingDisplacement2020}, however, ``there is a difference between research that is trying to identify gentrifying neighbourhoods and research that is trying to predict gentrification.'' In the context of EWS, this claim can be taken a step further: there is a difference, in turn, between attempts to \emph{predict gentrification} (or displacement) and a system that can productively serve the role of an \emph{warning system for gentrification}. In the latter case, it may be sufficient to predict a pre-determinant, or necessary condition for, these phenomenon. Such an EWS can then serve as a filter for further qualitative analysis \citep{yontoDevelopingGroundTruthingMultiScalar2020} or expert and community input.

With respect to model selection, EWS developers must determine, of the myriad potential methods available to them, how to develop a tool that actually makes predictions about future gentrification and/or displacement. In conventional studies of predictive modeling, this process involves three steps: 1) identifying a set of independent variables (or, synonymously here, \emph{features} or predictors), 2) identifying a (set of) potential algorithm(s) or statistical model(s) that use these variables to make predictions, and 3) "training" the model---finding the parameters that result in the best predictions--- using gathered data.   Early EWSs leveraged a variety of approaches to feature identification and model construction, most often via the use of expert input and standard statistical models \citep{chappleUseNeighborhoodEarly2016}. More recent work has responded to calls for "enhanced predictive analytics" \citep{greene2016if} that leverage \emph{machine learning} methods---statistical and/or numerical optimization techniques that prioritize predictive accuracy above all else---to make decisions about both which variables are predictive, and how to link them to the outcome \citep{readesUnderstandingUrbanGentrification2019}. 

Finally, once a predictive model has been developed, an evaluation of the model must be conducted. When performing evaluation of predictive models, care must be taken to ensure 1)  that evaluations are conducted that represent the true power of the model to \emph{predict} future outcomes, rather than the model's ability to \emph{explain} the present, 2) that the model is compared against adequate---and ideally simple---baseline models to determine whether the additional complexity of a predictive model is really necessary, and 3) that a number of metrics are computed to ensure that model predictions are useful and equitable.

The present work proposes and evaluates a new EWS for gentrification and displacement, created via a combination of new and old solutions to these challenges of data selection, operationalization, model selection, and model evaluation.  With respect to data selection, we leverage an increasingly popular option in the study of gentrification and displacement, namely parcel-level data on home transactions \citep{carlsonMeasuringDisplacementAssessing2021,dingGentrificationResidentialMobility2016}. We focus on publicly available data on home transactions; that is, when homes are bought and sold, and for what amount. These data are of interest because they are 1) both spatially and temporally granular, and 2) widely available to both researchers and city governments. 

With respect to operationalization, we argue that while data on home transactions do not directly capture the processes of gentrification or displacement \citep{carlsonMeasuringDisplacementAssessing2021,bartonExplorationImportanceStrategy2016}, there is good reason to believe that significant turnover in homeownership is a useful signal to predict in an EWS. However, with timestamped parcel-level data, one outstanding question is, \emph{what is the correct level of spatiotemporal aggregation at which we should, and can, make predictions} \citep{yontoDevelopingGroundTruthingMultiScalar2020}? Here, we explore various potential answers to this question using a novel experimental approach, providing empirical insight into the extent to which it is possible to make predictions about home transactions at different levels of granularity, and qualitiative insights into the potential utility of predictions at these various levels. 

With respect to model construction, we use a theory-driven approach to develop a broad range of features that we input into a modern machine learning algorithm. More specifically, we develop a set of features that are based on the notion of \emph{endogenous gentrification}--- roughly, that gentrification and displacement are spreading processes that can be predicted in a particular spatial region by considering recent indicators gentrification or displacement in nearby spatial regions. Finally, with respect to model selection, we evaluate the predictive power of our EWS via comparison of our model to a number of reasonable baseline methods, using a variety of evaluation metrics. 

Below, we describe our proposed EWS, our rationale for a case study of the system on Buffalo, NY, and a series of predictive experiments  that show the promises and pitfalls of our proposed system. Via a more detailed study-in-study examination of data from 2018, we also discuss how our findings relate to both the future of an EWS for displacement and/or gentrification in Buffalo. Equally as important, we discuss how future developers of EWSs might use our work as a guideline to better ensure that they produce models that have considered potential issues that arise during operationalization and model selection and evaluation.


\section{Background}



Recent work has provided cause for both caution and optimism for EWSs for gentrification and/or displacement. Optimistically, \citet{readesUnderstandingUrbanGentrification2019} build a predictive model that they argue can accurately forecast which neighborhoods will gentrify in London.  \citet{preisMappingGentrificationDisplacement2020}, among others, provide more cautionary tales, questioning the ability of EWSs to generalize beyond the narrow contexts in which they were created. And \citet{chappleUseNeighborhoodEarly2016} note that these systems have struggled to gain traction amongst critical stakeholders \citep{chappleUseNeighborhoodEarly2016}.

The present work focuses only on how one might develop an EWS that provides reliable predictions. It therefore does not explicitly address important questions about adoption.  Our efforts are relevant to this concern only in that accurate forecasts are, in many domains, a precursor to the adoption of predictive modeling. Because we focus on early warning systems, we also largely set aside broader discussions on gentrification and/or displacement, except where they are directly relevant to the creation of an EWS. For recent historical overviews that provide this context, we point the reader to \citet{lees2013gentrification}, on displacement to \citet{eastonMeasuringMappingDisplacement2020}, and on both to the work by \cite{zukGentrificationDisplacementRole2018}. 

There are also articles that provide a deeper overview on the historical development of EWS than we will touch on here. Specifically,  \citet{chappleUseNeighborhoodEarly2016} provide an expansive review on history of EWS for gentrification in the U.S., and then proceed identify and evaluate 11 current EWSs used in city governments.  \citet{eastonMeasuringMappingDisplacement2020} survey quantitative methods employed by academics to measure gentrification-induced displacement, and assess their connection to the theoretical underpinnings of the concept. Here, we review what these and other works help us to understand about how to address the four challenges with creating EWS noted above.

\subsection{Data Selection}

As noted above, most EWS in the U.S. rely exclusively on census data to provide forecasts. However, academics have increasingly sought to move beyond these data in their studies of gentrification and displacement. This includes the use of other "traditional" forms of data, such as surveys \cite{carlsonMeasuringDisplacementAssessing2021}, data on foreclosures \citep{williamsHomeForeclosuresEarly2013}, and home transactions \citep{yontoDevelopingGroundTruthingMultiScalar2020}. We also leverage home transaction data here, because such data have the advantage of being readily available and theoretically meaningful.

It is worth noting, however, that more innovative data sources have also shown to be useful in the analysis of gentrification and/or displacement. This includes "big data" sources such as Google Street view images  \citep{hwangDivergentPathwaysGentrification2014a} and social media data \citep{glaeserNowcastingGentrificationUsing2018,chappleMonitoringStreetsTweets2021}, as well as more difficult to gather administrative data on credit scores \citep{hwangUnequalDisplacementGentrification2020}. These data, in particular those from social media, are a rich source of important social and psychological precursors to and effects of gentrification and/or displacement \citep{bartonExplorationImportanceStrategy2016}, but also rich with myriad, difficult-to-measure biases \citep{olteanuSocialDataBiases2019,radfordTheoryTheoryOut2020}. While future work may incorporate these data productively into EWS, we therefore focus here on a more traditional dataset which harbors a more extablished and direct relationship to the phenomenon of interest.

\subsection{Operationalization}

Even before questions of operationalization, existing EWS differ in how they operationalize gentrification \citep{bartonExplorationImportanceStrategy2016,preisMappingGentrificationDisplacement2020} and/or displacement \citep{carlsonMeasuringDisplacementAssessing2021}.  Gentrification is generally agreed to be the process by which neighborhoods with low socioeconomic status experience increased investment and an influx of residents of higher socioeconomic status \citep{kolkoDeterminantsGentrification2007}. However, gentrification has also been identified by scholars as a “chaotic concept,” captured empirically in a number of different ways \citep{benediktsson2016taming, beauregard2013chaos,bartonExplorationImportanceStrategy2016}.  For example, EWSs have used continuously-valued indices \citep{readesUnderstandingUrbanGentrification2019} and home prices \citep{steif2017predicting}, as well as binary variables indicating whether or not a neighborhood has gentrified \citep{preisMappingGentrificationDisplacement2020}, as outcomes of interest. Similarly, displacement is widely understood to refer to  “an[y] involuntary residential move by a household”  \citep[][pg. 576, cf \cite{grier1980urban,vigdor2002does}]{carlsonMeasuringDisplacementAssessing2021}, but measurements of displacement vary in ways that create significant empirical disparities \citep{carlsonMeasuringDisplacementAssessing2021,eastonMeasuringMappingDisplacement2020}.

Closely aligned with the efforts presented here is the work of \citet{eastonMeasuringMappingDisplacement2020} compare three different proxy variables for displacement, referred to as the population approach, the individual-level approach, and the motivational approach. The population approach, most popular in the literature, looks at demographic changes, most often via an analysis of changes in low-income households. The individual-level approach tracks the number of households that report or can be identified as having moved. Finally, the motivational approach focuses on a subset of the individual-level approach, counting only households that moved involuntarily.  All three of these measures are constructed within a pre-specified spatiotemporal units, e.g. in census tracts over five year intervals. \citet{eastonMeasuringMappingDisplacement2020} compare these three different proxies both empirically, using survey and census data in New York City, and theoretically, arguing that the motivational approach ties most directly to the phenomenon of interest. They show, however, that the individual and motivational approaches are highly correlated. As discussed below, we use this strong correlation, in addition to other more theoretical arguments, to make the case that despite being a worse proxy for displacement, the individual-level approach is an appropriate and effective way to operationalize the outcome variable for an EWS.

What \citet{eastonMeasuringMappingDisplacement2020} does not consider, however, is the potential impact on correlations between these proxies as one varies the spatiotemporal granularity of the observation. Other EWS performance scholars have noted that census tract-level analyses can potentially miss the piecemeal nature of gentrification, in which harbingers of gentrification emerge at a smaller geospatial scale than the census tract, and/or in a smaller time span than census measurements allow study of \citep{eastonMeasuringMappingDisplacement2020}. Motivated by this, scholars have long looked at ways of operationalizing these phenomenon at smaller spatiotemporal scales \citep{yontoDevelopingGroundTruthingMultiScalar2020,eastonMeasuringMappingDisplacement2020,yontoDevelopingGroundTruthingMultiScalar2020}.  

Most notably, \citet{yontoDevelopingGroundTruthingMultiScalar2020} take an individual-level approach that computes a gentrification measure in a given spatio-temporal unit based on an index computed from parcel-level data. They compare these measures by evaluating their utility and validity according to domain experts, finding that density estimations of parcel-level data were found to be both accurate and useful. Their work emphasizes the importance of looking at ways of modeling and measuring gentrification and displacement that move beyond the granularities of census data, in particular to parcel-level data. Our work compliments their efforts with a more robust predictive experiment of how well \emph{changes} in aggregated parcel-level data can be predicted across various granularities of space and time.

Finally, it is critical to note that our work focuses on quantitiative operationalization, as opposed to more qualitative measures. \citet{bartonExplorationImportanceStrategy2016} emphasizes that qualitative work is critical for studying, and forecasting, a small number of nuanced cases of gentrification and/or displacement, whereas quantitative measures are more effective in settings wher the goal is to understand larger trends in an imperfect way. We argue here that the complimentary nature of quantitative and qualitative work can be seen in another way, where quantitative models help us to better understand where to look, or where \emph{not} to look, in large and complex problems like the identification of future gentrification and/or displacement \citep{mcfarlandSociologyEraBig2015}.




\subsection{Model Construction}

Many of the EWS used in practice are constructed by ingesting administrative data and then using domain expertise to construct rules or thresholds that define neighborhoods as being at risk for gentrification or not \citet{zukGentrificationDisplacementRole2018}.  Such models benefit from their simplicity and their ties to local knowledge, but can suffer in particular when attempts are made to generalize them across contexts \citep{preisMappingGentrificationDisplacement2020}. This issue of generalization is most apparent in the variables selected. \citet{preisMappingGentrificationDisplacement2020} find, for example, that across models used in four different U.S. cities for predicting gentrificadtion, ``of the 18 variables considered among the four models, only two variables appear in all four models.''  

Several more recent work have used traditional statistical modeling to address concerns about generalizability. This is because statistical modeling can, at least in theory, use general theoretical principles to derive a model, and then adapt to more local context by the parameters of the model when trained on data from, e.g., a specific city. For example, \citet{kolkoDeterminantsGentrification2007}  uses census-tract data from 1980-2000 to demonstrate that neighborhoods closer to city center and having older housing stock are most likely to gentrify, studying a number of U.S. cities with the same model. 

However, advanced computational methods have been put to use in a variety of ways to better understand gentrification and/or displacement. These efforts can be reasonably well bifurcated into two groups - those that use \emph{unsupervised} methods, and those that use \emph{supervised methods}.  Unsupervised methods are used to better understand datasets that have no clear outcome variable; examples include algorithms (including standard statistical approaches) to cluster data, and or work that reduces the dimensionality of complex phenomenon.  For example, \citet{delmelleMappingDNAUrban2016} use unsupervised methods to study neighborhood trajectories, including an analysis of gentrification. Their method includes a K-means clustering procedure that establishes discrete classes of neighborhoodsm and then a sequential pattern mining technique to the similarity of longitudinal sequences of evolution of neighborhoods. \citet{johnsonSmallAreaIndex2021} use principle component analysis, a dimensionality reduction technique, and a spatial smoothing model to create a small-area index of gentrification. Like the present work, part of their interest is in studying gentrification in small areas; however, their work intends to create an index of gentrification, rather than to create an early warning system.

Better connected to the present work, however, are \emph{supervised} machine learning approaches that try to predict--- or in our case, to forecast (predict into the future)--- an outcome relevant to gentrification and/or displacement. The most relevant example to the present work is that of \citet{readesUnderstandingUrbanGentrification2019}, who use machine learning to forecast gentrification using decennial, publicly available data on the city of London, UK. In their case, gentrification is operationalized using an index variable, and a model is trained on data from 2000-2010 to predict gentrifying neighborhoods from 2010-2020. The present work builds on these efforts, using machine learning to create an EWS in a similar spirit. However, our work has several distinct differences, in that we 1) study a much smaller, American city, 2) evaluate predictive accuracy along a much smaller and more varied number of spatiotemporal units of analysis, and 3) develop our model using a theory of endogenous gentrification, rather than a more general approach that uses myriad census variables.  In the sections below, we provide further comparisons to the efforts of \citet{readesUnderstandingUrbanGentrification2019} as well.

\subsection{Model Evaluation}

Evaluation of EWSs is a fraught, and sometimes ignored, process. As \citet[][pg.127]{chappleUseNeighborhoodEarly2016} put it, ``[f]ew developers have systematically assessed the validity of their gentrification and displacement predictions.'' Still, a number of recent articles have provided exemplary evaluations of both existing and new EWSs. 

\citet{preisMappingGentrificationDisplacement2020}, for example compare four gentrification and displacement risk models developed by and for the US cities of Seattle, Washington; Los Angeles, California; Portland, Oregon; Philadelphia, Pennsylvania, and implement all four methodologies to one city, Boston, Massachusetts. The authors then construct a comparative analysis, assessing which neighborhoods each model predicted to gentrify, and differences across models. 

More directly related is the work of \citet{eastonMeasuringMappingDisplacement2020}, who evaluate whether the three proxy variables for displacement they analyze provide comparable rank orders of neighborhoods based on their displacement magnitudes. We use a similar approach to evaluation with ranking metrics in our work, although adding additional metrics as needed. Both  \citet{bartonExplorationImportanceStrategy2016} and \citet{yontoDevelopingGroundTruthingMultiScalar2020} compare EWS predictions in some way with domain expterts, \citet{bartonExplorationImportanceStrategy2016} to articles from the New York Times, and \citet{yontoDevelopingGroundTruthingMultiScalar2020} to individuals in city government. These qualitative comparisons are of use in understanding the things that an EWS does, or does not, capture, and we here make similar comparisons to prior qualitative analyses of gentrification in Buffalo \citet{taylor2018buffalo}.

Finally, \citet{readesUnderstandingUrbanGentrification2019} provide a robust quantitative evaluation of their EWS for predicting gentrification in London. Our work follows theirs in the use of several baseline models and a variety of metrics to provide a more comprehensive comparison of the proposed EWS to other potentially more simple approaches.  We extend their efforts, however, by providing measures of error (e.g. standard deviations) for our outcome metrics, as well as providing an analysis of where and why our model err'ed. 

\section{Case Study - Buffalo, NY}

\begin{figure}[t]
    \centering
    \includegraphics[width=\textwidth]{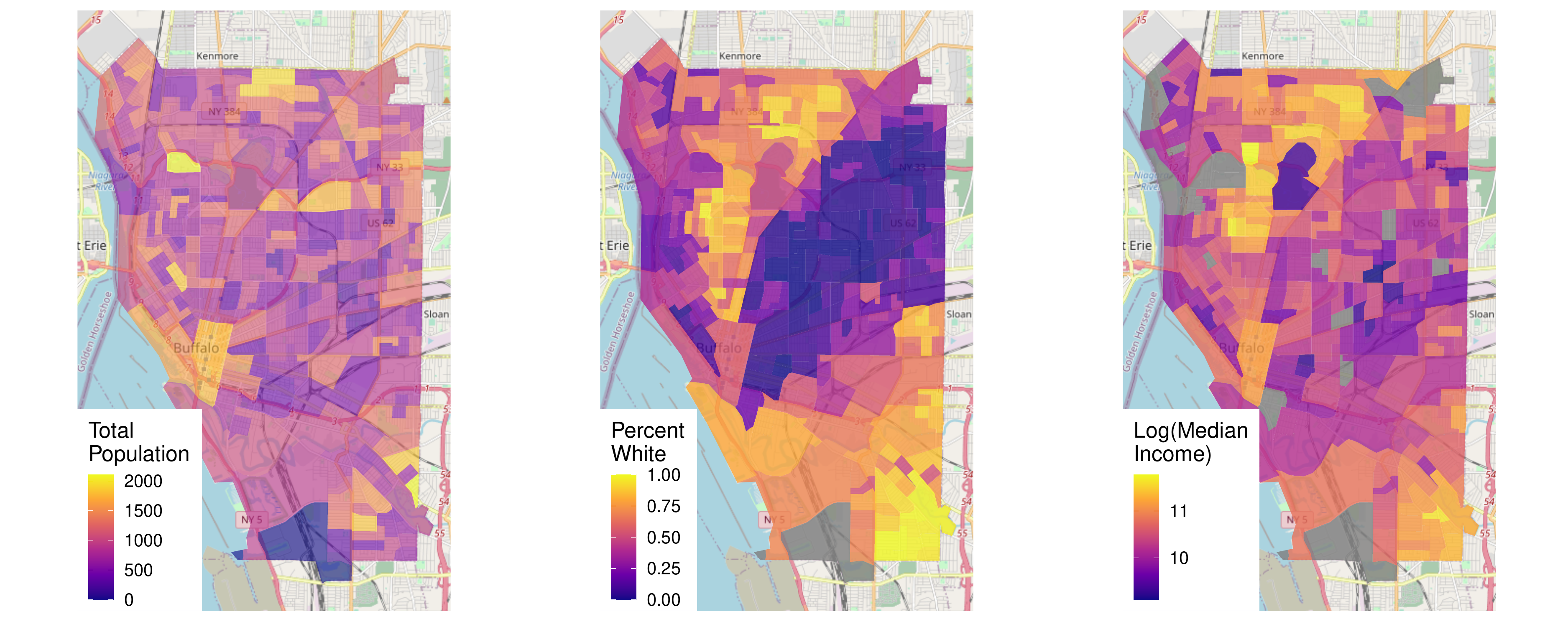}
    \caption{Data from the 2019 ACS five-year estimates, at the census tract level, for total population (left), percent of population identifying as White (center), and the logarithm of the median reported household income (right)}
    \label{fig:casestudy_fig}
\end{figure}

The city of Buffalo, New York, serves as the study region for this research. Located in Western New York State, Buffalo was one of the world’s most populous cities in the early 20th century. As with much of the Rust Belt, however, the city’s population and production steadily declined throughout the 20th century. In the wake of this decline came deteriorating conditions for low-income workers and an increase in racial segregation. Today, Buffalo is one of the nation's most racially segregated metropolitan areas, and the majority of its  impoverished residing in urban neighborhoods.  Following trends in the region, recent years have brought a steady return of new forms of business and industry to the area. The city is now seeing a significant house price increase, with prices increasing by 12\% by 2021, and the 2020 census showed that the population size of Erie county actually increased from a previous census for the first time in 70 years.\footnote{\url{https://www.wgrz.com/article/news/local/erie-county-population-grew-for-first-time-in-decades/71-84df3983-72c5-4fce-b804-2c727219ea37}.}  However, recent reports suggest this growth has largely eluded the city's Black residents.

Figure~\ref{fig:casestudy_fig} presents three maps that give an overview of the population distribution within the city of Buffalo. The city is often discussed as being composed of five parts, the borders of which are clearly discernible in the demographic makeup of the city presented in Figure~\ref{fig:casestudy_fig}. The West Side is a largely non-White region of the city that spans the northwestern edge of the city along the Niagara River. South of the West Side is Buffalo's Downtown, a small section of the city that is inclusive of federal and local government buildings, corporate officers, and the Buffalo Sabres' ice arena. To the east of Downtown is the South Side, a predominantly White Irish-Catholic neighborhood. North of the South Side is Buffalo's East Side, where most of the city's Black and low-income residents love.  The East Side is sandwiched between the South Side and, finally, the Elmwood/Delaware Park region of the city, where most of the affluent (and predominantly White) residents live.  This increase in investment and people coming to the city has created conditions favorable to gentrification, primarily in the West Side (coming from both Elmwood and in from the Niagara River) and on the East Side, moving up from Downtown. Using both qualitative and quantitative methods, scholars of the city have provided significant insight into the processes that underlie this gentrification \citep{taylor2018buffalo}.


\section{Methods}

Our analysis uses parcel-level transaction data from the city of Buffalo to construct an EWS. In Section~\ref{sec:data}, we describe our data selection, that is, how we collect the transaction data from publicly available data sources from the City of Buffalo and Erie County (where Buffalo is located). In Section~\ref{sec:outcome}, we describe how we operationalize the outcome for our EWS. We then discuss the model we develop for our EWS based on the theory of endogenous gentrification, and finally our evaluation process. 

\subsection{Data Selection}\label{sec:data}

Our primary data source begins with a set of 116,438 recorded transactions on 51,425 homes in the city of Buffalo, NY from January 1st, 2000 through December of 2019.\footnote{While data prior to the year 2000 is available, it is sometimes irregular. More specifically, pre-2000 data was not cleanly recorded in the system; most notably, certain years appeared to simply be missing. In addition, deed types were often missing or marked as a generic category that prevented us from being confident the transaction represented a sale. We also set aside data from after 2019, as we expect a potential discontinuity due to COVID. While the latter is of significant interest, it presents in our opinion a separate challenge for future study.} Our data collection process consists of four steps; as with the predictive modeling we describe below, all code used in our data collection is available publicly.  

First, we use a publicly accessible search engine provided by Erie County to construct a list of all homes within the county. To do so, we first use an open-access search page for Erie county to construct a list of the unique identifiers of every home in Erie county as of June, 2020. In New York state, this unique identifier is referred to as a \emph{SBL}, for "Section-Block-Lot". For each SBL, we then collect the information provided from the county on this search engine. This information includes identification of the home’s previous owners, including purchase and sale dates, the accompanying book page in the county’s deed book, and the buyer's name at the time. Additionally, it provides information on the property's value and tax payment history. This data is presented by the county in an HTML web page; to extract the information, we use the \texttt{BeautifulSoup} and \texttt{pandas} libraries in python.

Second, we use the deed information provided by the search engine to gather additional information about the price a home was purchased for. Again, we do so automatically, by constructing a tool to visit the appropriate page in the public records for Erie county. In addition to purchase price, the deed book also provides information on 1) the exact date of the transaction and 2) the type of deed recorded in the public record. The deed type is critical in differentiating home sales of interest to the present work from other events recorded in the public record, for example, name changes (which would otherwise appear like a change in ownership) or the receipt of a mortgage. Additionally, while the transaction date is also accessible on the county search engine, we found these dates to often be obviously incorrect. For example, dates such as 2088 were observed. Where dates disagree, we therefore use the information we drew from the deed itself, rather than the search engine.

Third, we geolocate each home to a latitude, longitude pairing. To geolocate the data, we first utilized a dataset from the city of Buffalo’s Open Data Portal\footnote{\url{https://data.buffalony.gov/}} which provides longitude and latitude coordinates for each parcel extracted from the county websites. These data were merged with the county data via shared SBLs. A small number of homes (98 homes, 0.7\% of all residential properties) did not have latitude/longitude information in the Buffalo Open Data portal, and were thus dropped from the study.

Finally, we perform filtering in order to focus on a subset of verifiable home transactions of interest for our study. 
The final stage of data collection involved filtering out data that was incomplete or not relevant to the present work.  First and foremost, the complete dataset covers transactions for various sorts of properties beyond residential homes. Given the focus on home transactions here, we exclude parcels marked as non-residential.  Second, as noted, not all transactions recorded by the county constitute a home sale. Home transactions were identifiable, however, based on the type of deed document listed in the deed book. Following a review of a range of deed document types, the decision was made to accept only a subset of document types (D1A, DEED, D1B, and D1BU) that we could verify were home transactions.  Transactions with these deed types made up 87.4\% of the remaining transactions, all others were dropped. Of the remaining 101,798 transactions, 26,638 (26.2\%) were for a cost of less than \$5. As part of our work considers the price of home transactions, and these transactions do not appear to reflect a  genuine price or value of a home, they were also removed.

In sum, then, the present study analyzes a dataset of all 75,160 home transactions captured by public records from 2000-2019 in the city of Buffalo, NY that are aligned with deed document types that indicate a home purchase.  As noted above, there are important advantages to working with this kind of home purchase data; namely, granularity and availability even in small cities like Buffalo. However, there are also important limitations to keep in mind. While many of these are discussed in the concluding section of this article, one important point to emphasize here is that the data does not contain information on renters. As such, the kind of displacement focused on here centers on how owners may be forced or desire to move, rather than the equally important (but more difficult to quantify) shift in renter behavior.

\subsection{Outcome Operationalization}\label{sec:outcome}

\begin{figure}[t]
    \centering
    \includegraphics[width=\textwidth]{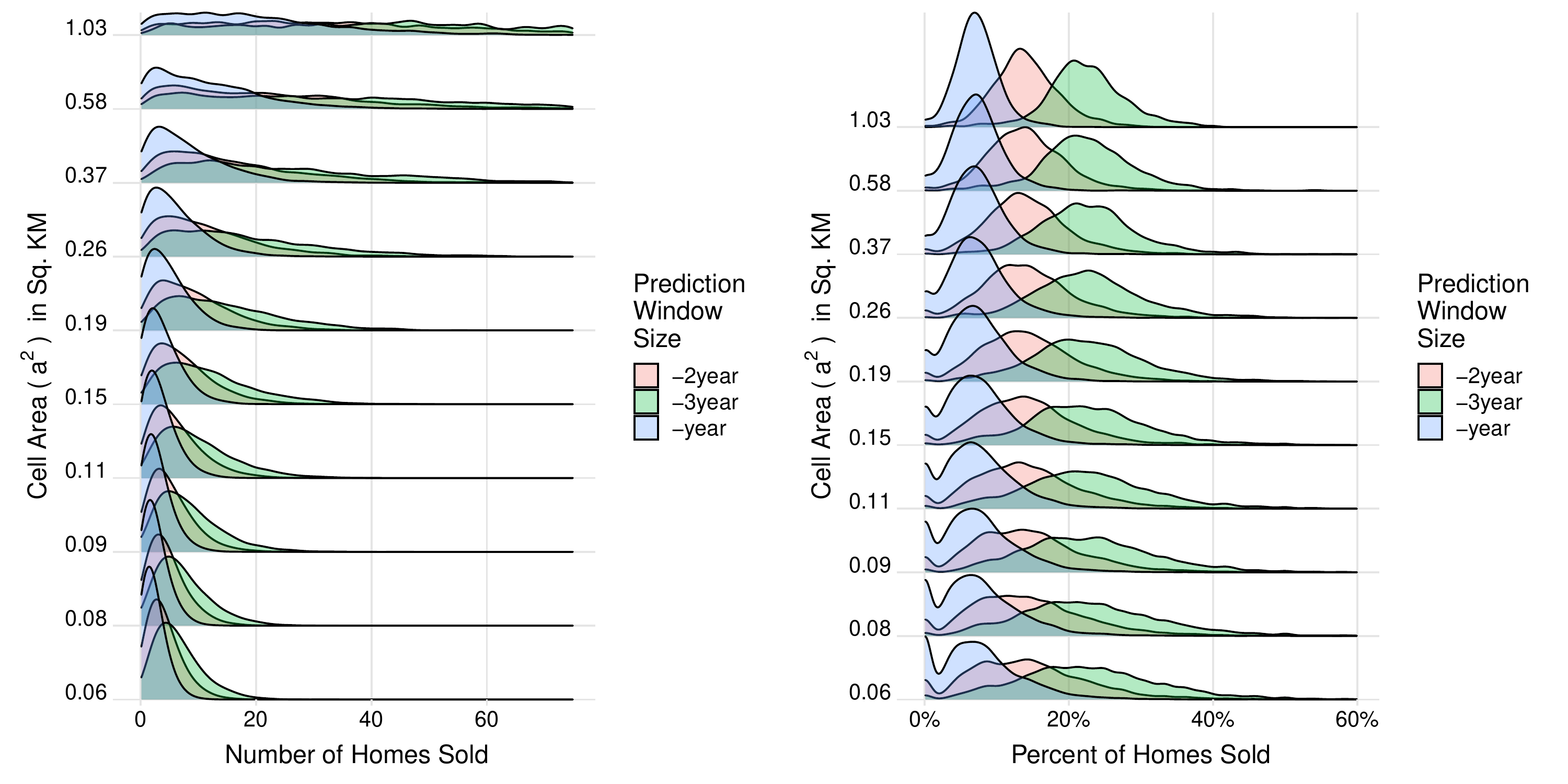}
    \caption{A) Left; For each grid cell size ($a^2$, x-axis) and prediction window size (1 year out, 2 years out, and 3 years out, different colored density displays), a kernel density plot representing the number of homes sold (y-axis) across all grid cells in all years studied. \hspace{\textwidth}
    B) Right, the same, except the y-axis is the outcome variable used in the study, namely the percent of homes sold in a grid cell }
    \label{fig:eda}
\end{figure}

As noted above, part of our effort is geared towards questions about the spatial and temporal resolutions under which it is possible to make predictions.  To this end, rather than focus on a specific geographic unit (e.g. the census tract), we opt to make predictions based on an $NxM$ rectangular grid. The use of a grid has the benefit of allowing for explicit modeling of spatial resolution and it's effect on predictive performance.  With respect to time, we evaluate predictions over a number of \emph{prediction window sizes}, ranging from one to three years.

The outcome variable our EWS seeks to predict therefore can be stated as the percentage of properties sold \emph{at least once} in a \emph{grid cell} with area $a^2$ during a specified time period $t$. More formally, given $a^2$ and $t$, we try to predict $y_{c,t}$, the percentage of homes bought in time period $t$ in grid cell $c$. In our prediction experiments, we vary both $a^2$ and $t$ to evaluate how predictive power varies as the spatio-temporal granularity of the grid overlaid on the city changes. Notably, we assume a fixed denominator; that is, we compute the number of home purchases in a grid cell, while fixing the total number of homes in the cell to be the number of homes that have ever been bought or sold in that cell as of June, 2020. This limitation is a function of our inability to track new builds; however, because new builds will appear as home transactions at the time of purchase, they will still be identified in the numerator of our outcome variable. Further, to ensure stability in our estimates of the proportion of homes transacted upon, we remove all grid cells with fewer than ten total homes. This results in a grid that does not necessarily capture the entirety of the city, see below for further discussion of this point.

Figure~\ref{fig:eda} provides summary data for grid cells using the full set of values of $a^2$ and $t$ we consider in this paper. Figure~\ref{fig:eda}a) shows that the number of homes sold varies drastically across spatial granularity, from a median of 3 when $a^2=3\,km^2$ (averaged across all prediction windows) to a median of 25 when $a^2 == 1.03$. In contrast, Figure~\ref{fig:eda}B) shows that the percent of homes sold in grid cells---our main dependent variable--- is relatively stable as spatial granularity increases. The median, for example, varies minimally, from 10.69\% to 10.81\%, across values of $a^2$ considered here. Further, there is a stable effect of temporal granularity, in that the difference caused by increased exposure time is relatively consistent across spatial granularity. In sum, Figure~\ref{fig:eda} therefore suggests that while the number of homes of course varies heavily across space and time, the (time-constant) \emph{rate} of home transactions, the outcome of interest here, may be relatively stable and thus more predictable.

Critically, however, we do \emph{not} claim that increased volume of home transactions \emph{is} displacement, nor is it a direct function of gentrification. With respect to the former, displacement is defined as involuntary movement; our outcome does not speak to whether or not moves are willful.  With respect to the latter, gentrification is a process that may precede or lag the displacement of long-standing residents.

However, there are four reasons why we believe that our outcome metric is appropriate \emph{for the task of developing an EWS for gentrification}.  First, while an increase in home transactions is not a sufficient condition for displacement and/or gentrification, it is by most accounts a necessary one. That is, displacement at least, by definition, \emph{requires} the movement of individuals out of and into a given area of a city.  Being able to predict this necessary signal of displacement can then allow for deeper study of a particular area, e.g. through interviews \citep{taylor2018buffalo} or surveys \citep{carlsonMeasuringDisplacementAssessing2021}, to determine whether or not gentrification is occurring.  Second, as noted above, there is a strong correlation between the number of overall home transactions in an area and displacement. \citet{carlsonMeasuringDisplacementAssessing2021} shows that at neighborhood level, the number of home transactions correlates reasonably well ($\rho$=.64), albeit with important differences, from a more granular, survey-based measure of displacement that also accounts for the fact that displacement aims to measure \emph{unwanted} moves.

Third, most studies on displacement seek to identify displacement due to gentrification \cite{eastonMeasuringMappingDisplacement2020}. However, focusing only on gentrification-induced displacement is problematic for two reasons. First, it requires that we have a single, concrete definition for a gentrifying neighborhood, and second, it can overlook the fact that even after neighborhoods have gentrified, they can still experience displacement at finer-grained spatial scales (e.g. within specific blocks) \cite{eastonMeasuringMappingDisplacement2020}. As such, our work does not seek to identify displacement only in neighborhoods that are candidates for gentrification, but again instead assumes that such decisions can be made after-the-fact by experts who are given EWS predictions.

\begin{figure}[t]
    \centering
    \includegraphics[width=.75\textwidth]{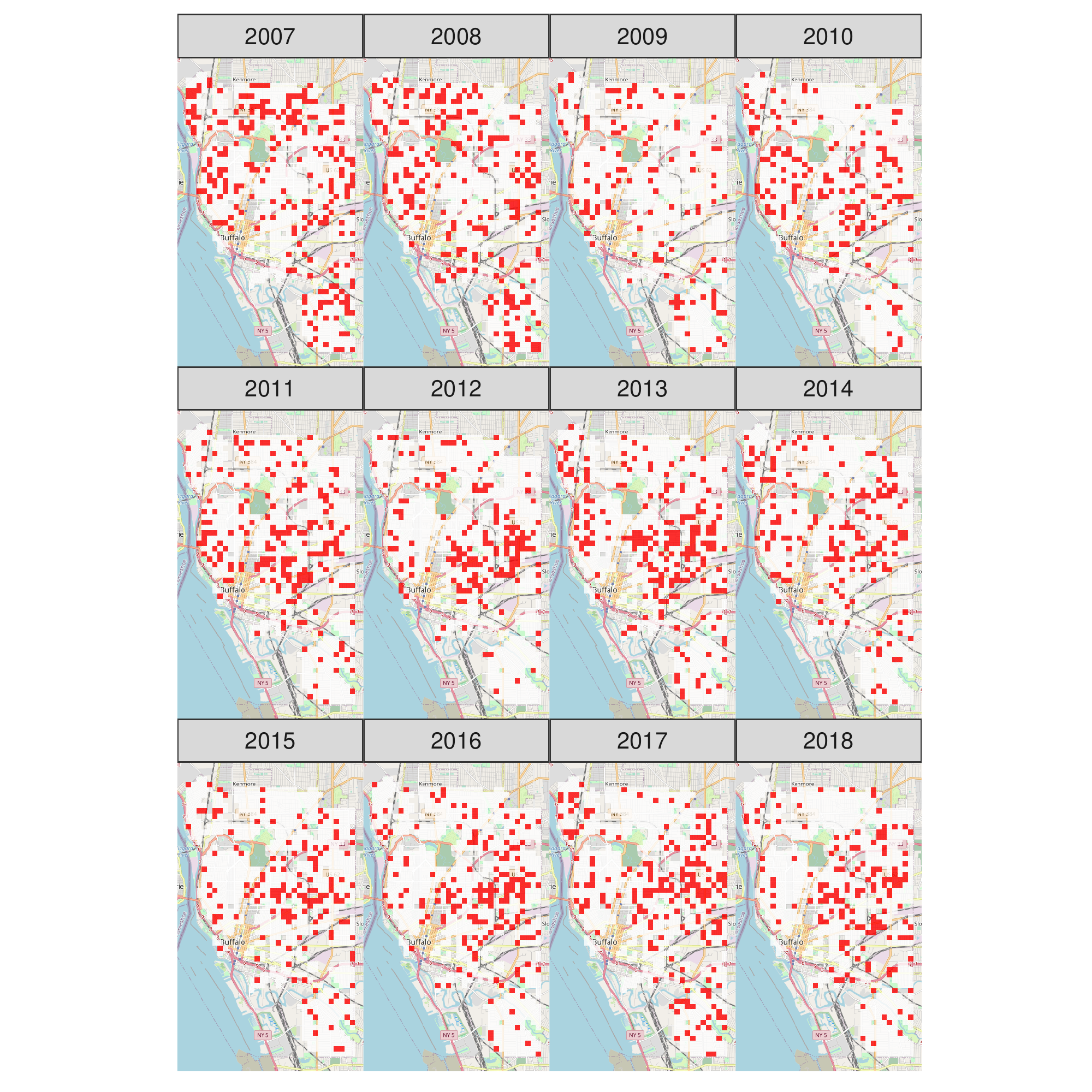}
    \caption{Each sub-figure represents a different year of transaction data, from 2011-2019.  Overlaid on a map of the city of Buffalo are white and red grid cells; areas with no grid cells have fewer than 10 homes and are thus excluded from our analysis. Cells are red if the percent of homes bought/sold in that cell was 1 standard deviation or higher above the mean value in that year.}
    \label{fig:grids}
\end{figure}

Finally, in addition to the theoretical arguments above, we find anecdotal evidence that our outcome metric is predictive of future gentrification. Figure~\ref{fig:grids} shows the areas with the highest rates of home transactions each year from 2011-2019. Compared with Figure~\ref{fig:casestudy_fig}, such grid cells lie predominantly in less White, lower income areas of the city, specifically on the city's West Side and East Side. This, in turn, corresponds with findings from the Turning the Corner Buffalo report discussed above, which focused on qualitative findings of gentrification in these areas.

\subsection{Model Construction}


The task of constructing a predictive model for our EWS itself has two steps. First, we define a set of independent variables, or \emph{features}, that we will use to inform predictions. Second, we define the model that we will consider to make predictions using those independent variables. 

\subsubsection{Feature Identification}\label{sec:features}


As with \citet{readesUnderstandingUrbanGentrification2019}, a complete description of the feature construction of our model is provided in the code and supplementary material released alongside this article. Here, we provide an overview that provides the intuition behind our modeling decisions. 

In order to create model features, we first aggregate our data according to particular values of $a^2$ and $p$. So, for instance, if $a^2 = 1\,km^2$ and $p = 1\,year$, the first step of our feature construction process is to aggregate our data into yearly (because $p=1\,year$) snapshots, each of which represents a set of grid cells with an area of 1 $km^2$. We would therefore have data for each grid cell for each year from 2000-2019. Here, we will refer to a spatial grid cell at a given point in time (e.g. in 2014)  \emph{entry}. 

For each entry, we compute three statistics: 1) the absolute number of transactions, 2) the percent of homes sold, and 3) the median home sale price. We then, for each of these statistics, compute a number of values that are used as features for the model. Here, as an example, we outline the features we create using one of these statistics, median home sale price. The process for creating features for the other two statistics is the same. 

The first features we use for each entry is simply the median home price for this entry in this time period. This feature encodes the notion that "the past is the best predictor of the future," but does not touch on the notion of endogenous gentrification.

To this end, we then add features for the average median home price for different \emph{spatial lags}.  That is, for each grid cell, we take the \emph{average} median home price of its surrounding cells, or \emph{neighbors}, and add this as a feature. Because there is no obvious answer to \emph{how many} nearby cells to include, we compute features for a range of assumptions. Specifically, we first compute the average median home price for the eight closest grid cells.  We then repeat this process for the closest 16, 24, 32, and 40 grid cells.

Capturing average values of surrounding cells encapsulates the idea of gentrification as a local spreading process. However, these averages can be biased by high values in individual nearby cells. As \citet{steif2017predicting} notes, we would also like to include features in our model that are indicative of a more homogenous trend towards (or away from) increasing home sales. A common approach to measuring this kind of spatial autocorrelation is via the use of a \emph{Local Indicator of Spatial Association (LISA)} coefficient. In our running example, when a grid cell's LISA value is high (low), the cell's median home price is more (less) similar to all of it's neighbors than we would expect by chance. In turn, high values might suggest an already gentrifying area, or one that has largely yet to be gentrified, while low values might indicate areas on the cusp of a wave of home buying and selling. 

For our EWS, we use the local Moran's I statistic as our measure of local spatial autocorrelation. Given this value for each entry, we then create a number of features, including the raw value of the statistic for each entry, the p-value of the statistic, and a four level categorical variable that denotes if the LISA statistic is above or below the mean, combined with whether or not the cell's median home price is above or below the mean. We also create features based on the spatial lag of LISA values, in particular the highest LISA value for a nearby cell (as above, within a given definition of "nearby"), and how far the cell is from this neighbor with the highest LISA value. Our last set of features for each entry seeks to capture where the entry is in relation to already gentrified and 2) potentially gentrifying neighborhoods. To this end, we include the distance to the cell with the highest median home price (or highest percent of homes with a transaction).

Finally, the features described to this point are based solely on statistics from time period $t$. However, the timescale over which the gentrification process unfolds is widely debated, and so statistics from previous time periods are also useful in the context of an EWS. Consequently, we also include as features for each entry all of the quantities described above, except in previous time periods as well. More formally, we compute the same features described here for time $t-1,t-2,.., t-\delta$. In our experiments here, we set $\delta = 3$. So, for example, to make predictions about 2018, we would include the features described here on data from 2017, but also 2016, and 2015 as well.

\subsubsection{Model}\label{sec:models}

Given the features we extract, our next task is to construct models that are able to use these features to make predictions. Following \citet{readesUnderstandingUrbanGentrification2019}, we use a \emph{random forest model}. Roughly, a random forest model makes predictions by randomly sampling the training data, and then constructing decision trees from those samples. To make a prediction, a random forest then uses a voting mechanism, where predictions each of the decision trees constructed on random subsets of the data are averaged.  We note that while other models were considered, we opted to include results only from the random forest here for three reasons. First, other models showed limited differences from the random forest. Second, given this, we find random forests are somewhat easier to explain concisely. Finally, using them follows prior precidence on EWS.

\subsection{Model Evaluation}\label{sec:evaluation}

Evaluating predictive models requires that one separate the data used to infer model parameters from the data used to evaluate model predictions. Not separating these can lead to what is called \emph{overfitting}, where one generates estimates of model performance that are overly optimistic because the model has ``memorized" the data it is given.  The most common approach to avoiding overfitting is a practice entitled \emph{cross-validation}, where the data is repeatedly split into a \emph{training} dataset, used to infer model parameters, and a \emph{testing} dataset used to test the model.

For predictions over time, a special kind of cross-validation, known as \emph{temporal cross validation}, is needed in order to ensure that the model is evaluated according to its ability to predict the future from the past. In temporal cross-validation, we first pick a \emph{train/test split date}, which defines the date at which we will try to make predictions into the future, given all data prior to that date. So, for example, if the train/test split date was January 1st, 2018, we would try to make predictions on home transactions after January 1st, 2018, given all data prior to that date. We then vary the train/test split date several times to ensure our results are not dependent on the use of a single split point. 

In our experiments, we evaluate the predictive performance of our model using temporal cross-validation across the suite of spatial and temporal resolutions. For temporal resolutions, we aim to make predictions over prediction windows ($t$) of one year, and two years, and three years into the future. For spatial resolutions, we vary $a^2$ from 0.06 $km^2$ to 1.03 $km^2$. For reference, the average census tract in Buffalo measures 0.37 $km^2$. Predictions were made by jumping forward one year across all cases, starting in 2009. So, for example, for a prediction window $t$ of two years, we would first predict the proportion of homes sold at least once in all grid cells in 2009-2011, and then 2010-2012, and so on.

Once obtaining predictions from our model, two additional steps are required. First, we must identify a set of \emph{baseline} models, or simple but potentially powerful predictive models to which we can compare our more involved approach. Second, we require a set of metrics, or quantities, that can be used to state the quality of each model.  We describe each of these in turn in the following subsections.

\subsubsection{Baseline Models}

We use three baseline models here. The first, which we call the \emph{Previous Year, Single Cell} model, predicts $y_{c,t}$ to be the value from the previous time period $t-1$, i.e. $y_{c,t-1}$. The second model, which we call the \emph{Previous Year City Average} does the same, except instead of using the previous value for a single grid cell, it uses the average over all cells, $\widetilde{y_{*,t-1}}$. For example, if the time span used were to be a year, this model would predict, for each cell, the average percentage of transactions across all of Buffalo in the previous year.   Finally, the third model, which we call the \emph{All Years, Single Cell} model, predicts $y_{c,t}$ to be the average in that same cell across all years, i.e. $\widetilde{y_{c,*}}$.

Baseline models are useful for two reasons. The first is to ensure that the prediction task is non-trivial; if simple models can make near-perfect predictions, then there is little reason to concern oneself with more complex approaches. Second, these baseline models provide an important comparison point in non-trivial tasks. If more complex models do not exceed baselines, then one must be careful in interpreting outputs from more advanced models. This is because the advanced model is likely simply picking up on correlates of the more simple features used by the baselines. In contrast, to the extent that more complex models do outperform baseline models, one can have confidence that features leveraged by more complex models do provide valid signal above and beyond simple quantities used by the baselines.

\subsubsection{Evaluation Metrics}

Metrics selected for the evaluation of an EWS should reflect the variety of ways in which such a system might be used. To this end, we here select four traditional evaluation metrics that can be used to quantify the extent to which model predictions match real data. Following \citet{readesUnderstandingUrbanGentrification2019}, we first use the \emph{root mean squared error (RMSE)}, defined mathematically as:

\begin{equation}
\begin{aligned}
RMSE  &= \sqrt{\dfrac{\sum_{c} (y_c-\hat{y_c})^2 }{N}}  \\
\label{eq:litdiff}
\end{aligned}
\end{equation}   

RMSE is useful in providing an estimate of the extent to which the exact predicted values of the model match the data.  

However, most EWS are not concerned with exact predictions, but rather in \emph{ranking} the objects of interest relative to each other \citep{bartonExplorationImportanceStrategy2016,eastonMeasuringMappingDisplacement2020}. This is because EWS are most often used as tools to narrow the focus of more extensive, and often more expensive, investigative work that often relies on domain expertise and/or further qualitative and quantitative analysis. Because of this, we include two additional evaluation metrics. First, we use a traditional statistic, Kendall's Tau, a rank-order correlation metric defined as:

\begin{equation}
\begin{aligned}
\textit{Kendall Tau-b Correlation}  &= \dfrac{P-Q}{\sqrt{(P+Q+X_0)(P+Q+Y_0)}}  
\label{eq:litdiff}
\end{aligned}
\end{equation}   
Second, drawing from the literature on recommender systems in comptuer science, we measure the \emph{Normalized Discounted Cumulative Gain (NDCG)}, a rank metric that weights the top of the ranking more heavily.

\subsection{Case Study}

In addition to presenting results on the predictive capabilities of the model across various years and spatiotemporal granularities, we find it useful to also present results from a single instantiation of the model to further demonstrate the model's utility and potential drawbacks. To this end, we use the model to make predictions about home transactions with a fixed grid cell size $a^2= 0.145\,km^2$ (approximately half the size of a census tract) and temporal resolution ($t= 1\,year$) for a given year (2018), and explore those predictions in various ways.

\section{Results}

\subsection{Main Findings}

\begin{figure}[t]
    \centering
    \includegraphics[width=\textwidth]{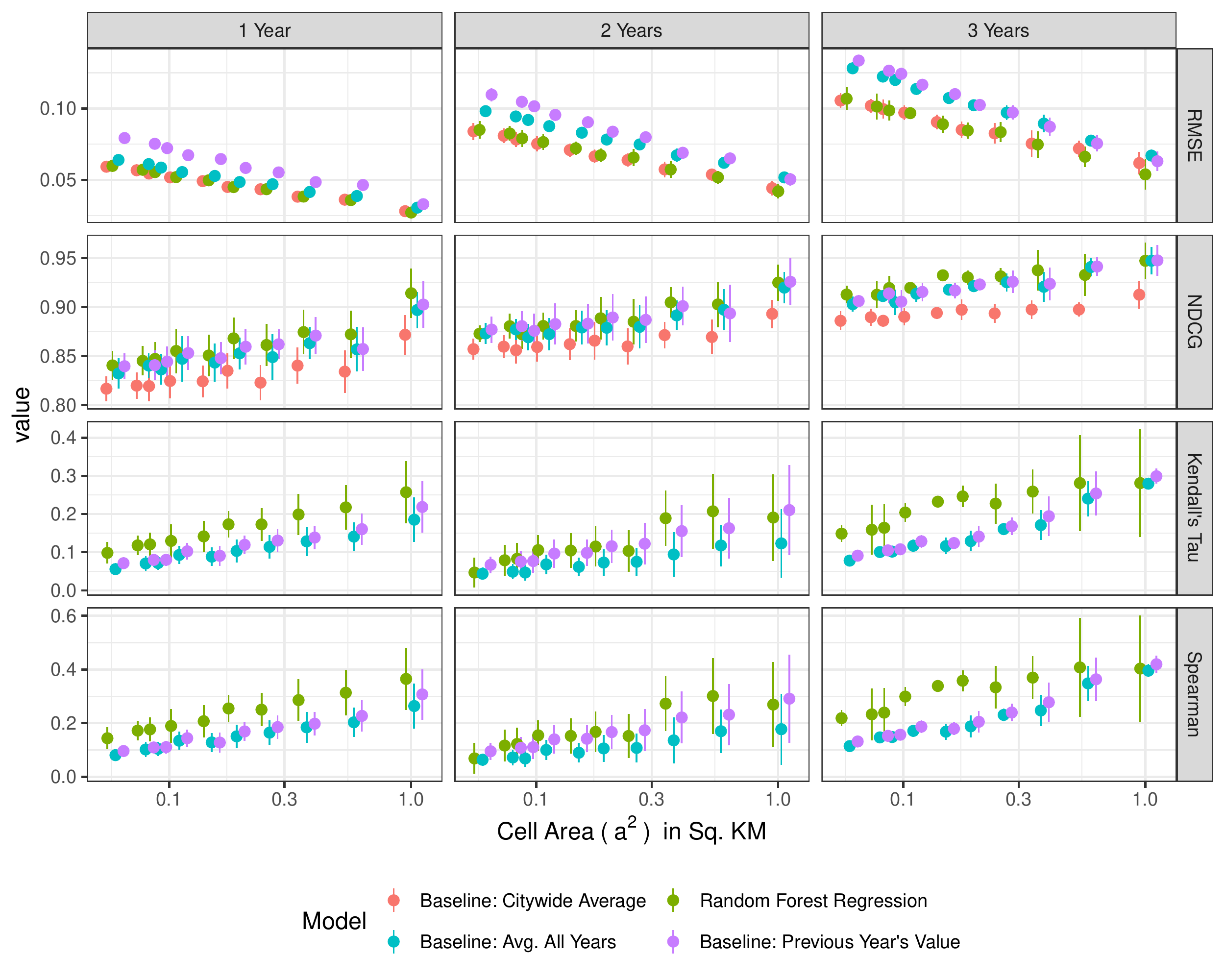}
    \caption{Results on the three evaluation metrics (different rows of subplots) for our EWS and the three baseline models (different color point ranges) for each grid cell size (x-axis) and prediction window size (different columns of subplots). Error bars are across the different folds of the temporal cross-validation, using normal 95\% CIs.}
    \label{fig:highlevel}
\end{figure}

Figure~\ref{fig:highlevel} shows that variations across our predictive models are relatively consistent across grid cell size and window size variations, with a log-linear growth in error across all metrics we consider as spatiotemporal granularity becomes more refined, and a slight but constant increase in predictive power as temporal granularity increases.  Notably, the relative change in error is largest for RMSE; this reflects in part an overall increase in the magnitude of the outcome (and thus, trivially, an increase in the differences possible between the outcome and the prediction).  In general, then, our EWS, as well as the baselines we model, are more predictive at larger spatial granularities, but the magnitude of error changes are limited once we move beyond the largest cell sizes. More specifically, we see that beyond the mean census tract size (approximately 0.37), prediction error shows limited relative changes, indicating the possibility of making predictions at finer spatial granularities without a significant decrease in model error. The model also shows slightly better predictive power at 2-3 years out than it does at 1 year out, reflecting the potential for predicting longer-term changes more accurately than shorter term changes.

\begin{figure}[t]
    \centering
    \includegraphics[width=\textwidth]{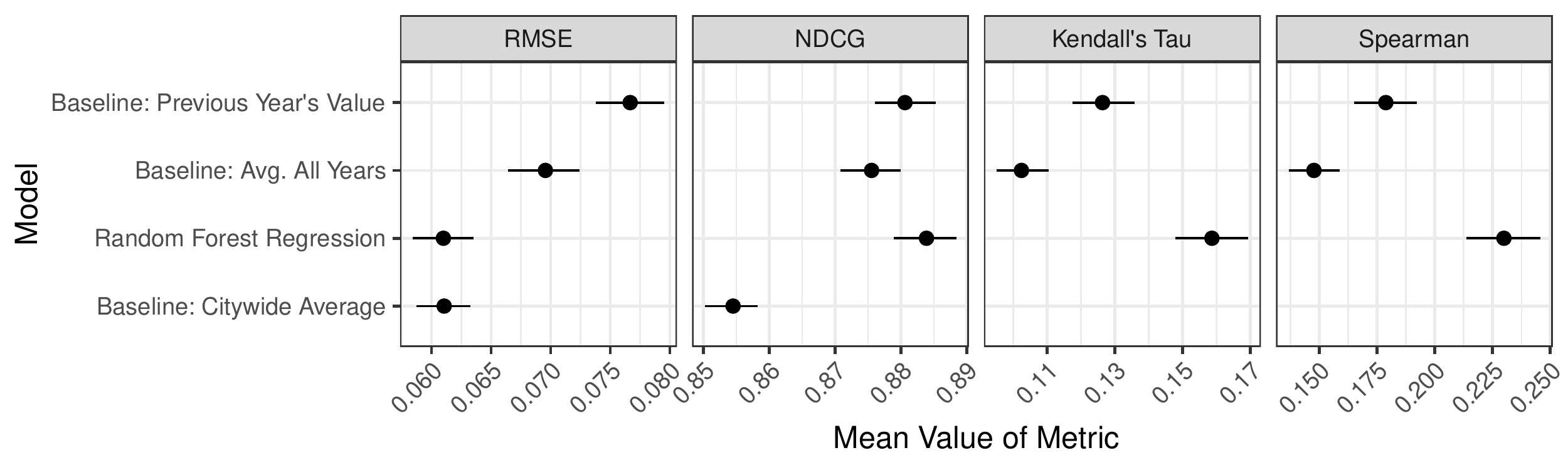}
    \caption{Results for our model and the three baselines (y-axis) for each of the three outcome metrics (different sub plots) averaged across all grid cell and prediction window sizes.
}
    \label{fig:mod_avg}
\end{figure}

Equally as important in Figure~\ref{fig:highlevel} is that our EWS slightly out-performs the baseline models on some, but not all, metrics, and that these model performance differences are relatively stable across spatiotemporal granularity. Because of this relative stability, it is reasonable to assess overall differences in model performance, averaging over these different parameters of $a^2$ and $t$ in our experiment. Figure~\ref{fig:mod_avg} displays these results; the primary takeaway is that while our EWS does not show a statistically significant improvement over all baselines on any one metric, it is the only model to perform as good or better than all other models on each of the different metrics.

In sum, we present two main findings. First, surprisingly, that our ability to predict our outcome of interest only marginally decreases as we increase the spatiotemporal granularity of our observations. There are several potential reasons for this. First, and most practically, increased spatiotemporal granularity provides additional training data that can be used to refine model predictions. Second, as noted above, increased spatiotemporal granularity creates increased variance in the outcome variable, which may also allow for an easier prediction task. Finally, it is possible that the process of interest, clustered home transactions, occurs on a relatively fine-grained spatiotemporal scale, and as such is genuinely predictable at finer grains. Further work is, however, necessary to address which of these explanations drives our observations.

Second, we find small but consistent performance gains in our endogenous gentrification-based EWS over reasonable baseline models.  This can be interpreted in two ways. First, the performance gains indicate that a more complex model is indeed useful in identifying future home transactions.  Second, however, is that these gains are, on any given metric, not appreciably different from very simple approaches. Given the dearth of comparisons to such simple baselines in the existent literature on EWS, we therefore find it important to emphasize the tradeoffs between more complex approaches and those that make simple but intelligent use of existent data.

\subsection{2018 Case Study}

\begin{table}[t] \centering 
\small
\begin{tabular}{@{\extracolsep{5pt}}lcc} 
\\[-1.8ex]\hline 
\hline \\[-1.8ex] 
 & \multicolumn{2}{c}{\textit{Dependent variable:}} \\ 
\cline{2-3} 
\\[-1.8ex] & \% Homes Sold & Prediction Error \\ 
\\[-1.8ex] & (1) & (2)\\ 
\hline \\[-1.8ex] 
\% Black residents & 0.015$^{***}$ & 0.0001 \\ 
  & (0.005) & (0.001) \\ 
  & & \\ 
\% Hispanic Residents & $-$0.002 & $-$0.006$^{***}$ \\ 
  & (0.004) & (0.001) \\ 
  & & \\ 
Log(Mean Income) & $-$0.015$^{***}$ & $-$0.010$^{***}$ \\ 
  & (0.004) & (0.001) \\ 
  & & \\ 
\% Homes Sold &  & $-$0.105$^{***}$ \\ 
  &  & (0.001) \\ 
  & & \\ 
 Constant & 0.076$^{***}$ & $-$0.013$^{***}$ \\ 
  & (0.005) & (0.0003) \\ 
  & & \\ 
\hline \\[-1.8ex] 
Observations & 851 & 851 \\ 
Log Likelihood & 1,310.514 & 2,817.103 \\ 
$\sigma^{2}$ & 0.003 & 0.0001 \\ 
Akaike Inf. Crit. & $-$2,609.027 & $-$5,620.206 \\ 
Wald Test (df = 1) & 7.404$^{***}$ & 7.395$^{***}$ \\ 
LR Test (df = 1) & 8.208$^{***}$ & 7.014$^{***}$ \\ 
\hline 
\hline \\[-1.8ex] 
\textit{Note:}  & \multicolumn{2}{r}{$^{*}$p$<$0.1; $^{**}$p$<$0.05; $^{***}$p$<$0.01} \\ 
\end{tabular}
  \caption{Regression Results for Two Linear Spatial Simultaneous Autoregressive Models. Observations for both models are cell grids where $a^2=X$, $t=$1 year for the year 2018. All independent variables (rows of the matrix) except for \% Homes Sold are drawn from the 2018 ACS 5-year data and interpolated from block groups to grid cells. The dependent variable of the first model (first column) is our primary outcome variable (\% Homes Sold), for the second (second column), it is the difference between the primary outcome and the prediction of our EWS. All indepedent variables are centered and scaled  by two standard deviations.} 
  \label{tab:reg_results} 
\end{table}

Table~\ref{tab:reg_results} presents results from two regression models. The first seeks to predict the dependent variable of our EWS from three interpolated census variables: the percent of Black residents in a grid cell, the percent of Hispanic residents, and the logarithm of the mean income of households in the cell.  Results show that a one standard deviation increase in the percent of Black residents in a grid cell is associated with a 1.5\% increase in home cells, and a one standard deviation decrease in logged mean income is associated with a 1.5\% increas as well. These results further suggest that while our outcome metric is not predefined to be be centered on areas ripe for gentrification, it is associated with typical predictors of gentrification and/or displacement.

The second regression model in Table~\ref{tab:reg_results} seeks to predict the prediction error of our model, that is, the difference between our prediction and the true outcome. We find that this error is also significantly associated with census variables: a one standard deviation increase in percent Hispanic residents results in a slight (.6\%) but statistically significant deviation of the model's prediction from the true outcome, such that the model is slightly more likely to under-predict homes bought/sold in Hispanic neighborhoods. A similar finding is observed with income - lower income neighborhoods are associated with a prediction that underestimates home purchases by 1\%, controlling for other factors. Finally, there is a much larger effect of the dependent variable itself - a one standard deviation increase in percent homes sold in an area is associated with an under-prediction of 10.5\% on average, controlling for the other factors in the model.

The results in this second regression model in Table~\ref{tab:reg_results} suggest two important points. First, blind use of an EWS may result in increasing inequality - if the model underpredicts potential gentrification in lower income and/or Hispanic neighborhoods, users of the model may be less inclined to allocate resources to these places. Second, it is important to be cautious about using the absolute values of model predictions, because they may not be well-calibrated; that is, as with many predictive models, they may struggle to predict extreme values and thus suffer from a form of regression to the mean. This is a potentially non-obvious but nonetheless important point; for example, \citet{readesUnderstandingUrbanGentrification2019} use their similar model to forecast slowing gentrification in London, but do not consider the fact, as far as we are aware, that this finding could entirely be caused by this under-prediction phenomenon.

\begin{figure}[t]
    \centering
    \includegraphics[width=\textwidth]{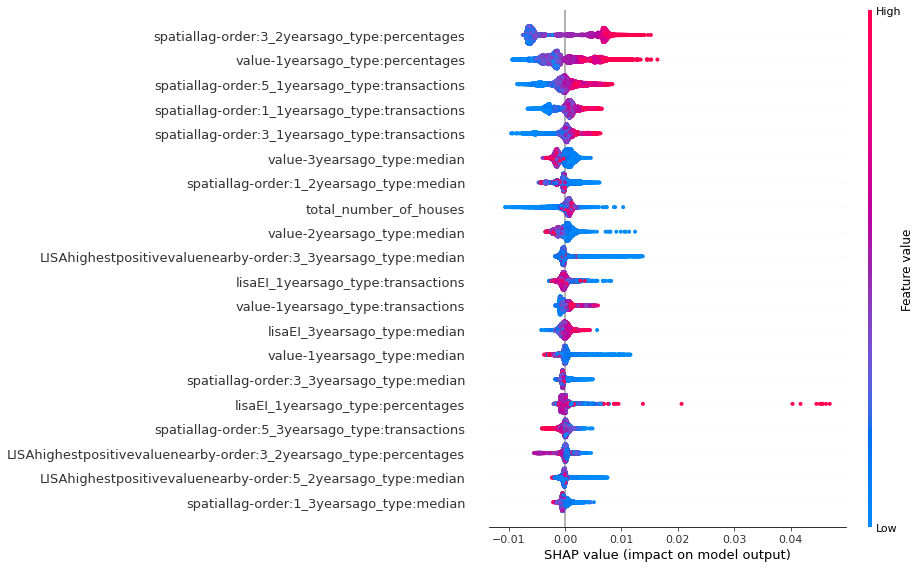}
    \caption{Variable importance, computed using SHAP, for an EWS for 2018 with $a^2= X$ and $t=$ 1 year. SHAP values (y-axis) are given for the full training data  (each dot) for the 25 most important independent variables in terms of mean absolute SHAP values (x-axis). Color represents the magnitude of the feature value; all features are rescaled from 0 (blue, lowest value) to 1 (orange,highest value).}
    \label{fig:shap}
\end{figure}

Despite these limitations of the predictive models, one practical use for them beyond the baselines presented is that we can explore features of grid cells that mare associated with future increases in home sales. Figure~\ref{fig:shap} displays the top 25 most predictive features in the model using a common measure of variable importance in random forest models, the  \emph{SHapley Additive exPlanation (SHAP) values} \citep{lundberg2017unified}. SHAP values can be computed for each independent variable for each observation (here, a grid cell) for a predictive model. The SHAP value for that variable, for that grid cell, then can be understood as the expected difference in the outcome (\% homes sold) given that value of the independent variable in that grid cell.  SHAP values can therefore help us understand how a given variable in a grid cell is associated with a change in the outcome.

The variables in Figure~\ref{fig:shap} present evidence that features in our model that are representative of endogenous gentrification are useful in forecasting gentrification. In particular, the most important feature to our model, in terms of mean absolute SHAP value, is the average percent of homes sold in the nearest 15 neighbors of a grid cell two years prior to the current year.  Critically, this feature is more important than features that are indicative of the grid cell itself, including how many homes are in the cell, and the previous value for percent homes sold. 

\begin{figure}[t]
    \centering
    \includegraphics[width=\textwidth]{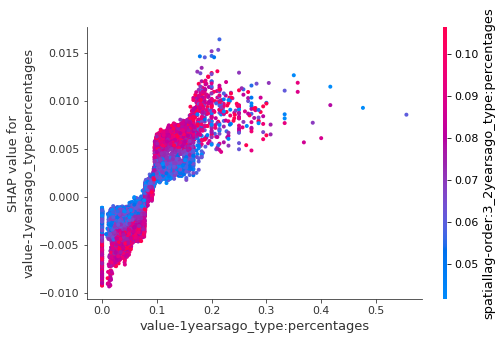}
    \caption{Caption}
    \label{fig:shap_interact}
\end{figure}

However, a feature of more complex modeling approaches, and in particular tree-based models, is that we can move beyond individual predictors to better understand how sets of predictors interact. Figure~\ref{fig:shap_interact}, for example, plots the two most important predictors in our model as determined by their mean absolute SHAP value, and shows a non-linear relationship between these two variables and the outcome. More specifically, our model suggests that in cells where \emph{greater} than 10\% of homes were sold in 2015, increased home sales in nearby grid cells in 2016 were associated with a \emph{further increase} in home transactions in 2017. In contrast, when \emph{fewer} than 10\% of homes in a cell were sold in 2016, increased home sales in nearby cells was associated with \emph{fewer} future transactions.

This outcome from our model suggests two points. First, it provides questions for future study about how endogenous gentrification may not only be a spreading process, but also a compounding one, in which a "rich get richer" effect occurs as displacement in one area exacerbates nearby displacement, which in turn creates further displacement, and so on. Second, it similarly suggests the converse; namely that locations which are insulated from gentrification cluster together.

\section{Conclusion}

We construct and evaluate a novel early warning system for gentrification based on data that is readily available to many U.S. cities. We show potential benefits and drawbacks of such a system; for example, we show that such a model can more accurately forecast future signals of gentrification across a number of spatial and temporal levels of aggregation, but that it may do so at the cost of furthering the inequalities produced by these processes in the first place.

As such, we argue that quantitative tools such as the one presented here \emph{must be used in close collaboration with qualitative methods}, e.g. those developed and conducted by \citet{taylor2018buffalo} and \citet{silverman2019there} for the city of Buffalo. Put another way, the tool we created is unusable without additional insights. Future work on early warning systems would, however, benefit from playing to the strengths and weaknesses of available technology, expertise, and lived experience. We see our work as a critical element in the advancement of this line of research and practice.

\bibliographystyle{agsm}
\bibliography{sample}

\end{document}